\documentclass[%
 reprint, 
 amsmath,amssymb,
 aps,
 nofootinbib
 prb,
 superscriptaddress
 ]{revtex4-1}
\usepackage{scrextend}
\usepackage{graphicx}% Include figure files
\usepackage{dcolumn}% Align table columns on decimal point
\usepackage{bm}% bold math
\usepackage{hhline}
\usepackage{tabularx}
\usepackage[utf8]{inputenc}
\usepackage[english]{babel}
\usepackage[bottom]{footmisc}

\begin{document}

\preprint{APS/123-QED}

\title{
Polytypism at its best: improved thermoelectric performance in Li based Nowotony-Juza phases
}
\author{Uday~Chopra}\thanks{U.C. and M.Z. contributed equally to this work.} 
\affiliation{Indian Institute of Technology Roorkee, Department of Chemistry, Roorkee 247667, Uttarakhand, India}
\author{Mohd~Zeeshan}\thanks{U.C. and M.Z. contributed equally to this work.}
\affiliation{Indian Institute of Technology Roorkee, Department of Chemistry, Roorkee 247667, Uttarakhand, India}
\author{Shambhawi~Pandey}
\affiliation{Indian Institute of Technology Roorkee, Department of Chemistry, Roorkee 247667, Uttarakhand, India}
\author{Harish~K.~Singh}
\affiliation{TU Darmstadt, Theory of Magnetic Materials, Department of Materials- and Earth Science, Alarich-Weiss-Str.~16, 64287 Darmstadt, Germany}
\author{Jeroen~van~den~Brink}
\affiliation{Institute for Theoretical Solid State Physics, IFW Dresden, Helmholtzstrasse 20, 01069 Dresden, Germany}
\author{Hem~C.~Kandpal}\email{Corresponding author: hem12fcy[at]iitr.ac.in}
\affiliation{Indian Institute of Technology Roorkee, Department of Chemistry, Roorkee 247667, Uttarakhand, India}

\date{\today}% It is always \today, today,

\begin{abstract} 
In principle thermoelectricity is a viable route of converting waste heat into electricity, but the commercialization of the technology is limited by its present efficiency. In quest of improved materials, utilizing an \textit{ab initio} approach, we report polytypism induced improved thermoelectric performance in Li based Nowotny-Juza phases LiZn\textit{X} (\textit{X} = N, P, As, Sb, and Bi). In addition to LiZnSb, we find that cubic LiZnBi is energetically more favorable than the hitherto explored hexagonal phase whereas the hexagonal polytypes of cubic LiZnP, and LiZnAs are likely to be stabilized by pressure -- hydrostatic pressure can be aided by internal pressure to facilitate the phase transition. We find a pronounced impact of the polytypism on thermoelectric properties. We determine conservative estimates of the figure of merit and find that while power factor and figure of merit values are improved in hexagonal phases, the values in cubic phases are still excellent. The \textit{ZT} values of cubic and hexagonal LiZnSb at 700 K are 1.27 and 1.95, respectively. Other promising \textit{ZT} values at 700 K are 1.96 and 1.49 of hexagonal LiZnP and LiZnAs, respectively, for \textit{n}-type doping. Overall, our findings proffers that the Nowotny-Juza phases are a new potential class of thermoelectric materials. 
\begin{description}
\item[Usage]
Secondary publications and information retrieval purposes.
\item[PACS numbers]
May be entered using the \verb+\pacs{#1}+ command.
\item[Structure]
You may use the \texttt{description} environment to structure your abstract;
use the optional argument of the \verb+\item+ command to give the category of each item. 
\end{description}
\end{abstract}

\pacs{Valid PACS appear here}% PACS, the Physics and Astronomy
\maketitle

%\tableofcontents

\section{Introduction}
The thermoelectric effect offers the potential to convert environmental waste heat into electricity \cite{Samanta17, Zhao17}. However, the commercialization of thermoelectric materials is limited since their efficiency is governed by conflicting parameters such as Seebeck coefficient, \textit{S}, electrical conductivity, \textit{$\sigma$}, and thermal conductivity, \textit{$\kappa$}, by the relation \textit{ZT} = \textit{S$^2\sigma$T/$\kappa$} (\textit{$\kappa$} = \textit{$\kappa_e$} + \textit{$\kappa_l$}) \cite{Boona17, He17}. In recent years, half Heusler alloys, \textit{XYZ}, have gained much attention owing to substantial thermoelectric properties comparable to conventional Bi$_2$Te$_3$ and PbTe based materials \cite{Fang17, Yan12, Yu09, Sakurada05}. 

Nowotny-Juza phases are a special derivative of half Heusler alloys when \textit{X} belongs to the Group-I, \textit{Y} to the Group-II, and \textit{Z} to the pnictogen family \cite{Wollmann17}. The crystal structure of Nowotny-Juza phases, \textit{A$^I$B$^{II}$C$^V$}, comprises a hypothetical zincblende type lattice of [\textit{BC}]$^-$ interpenetrated by a face-centered lattice of \textit{A}$^+$ \cite{Kuriyama99, Kuriyama88, Kuriyama87, Kuriyama91, Kuriyama96, Kuriyama94}. The zincblende type covalent framework offers the high mobility of charge carries whereas the filled octahedral sites (\textit{A}$^+$) provides ionicity to the system, thereby reducing the thermal conductivity. The filled octahedral sites may also be helpful in scattering the phonons by acting as rattlers \cite{Kishimoto08}. This proffers that Nowotny-Juza phases could be promising candidates for thermoelectric applications.
  
Within the Nowotny-Juza phases, the LiZn\textit{X} (\textit{X} = N, P, As, Sb, and Bi) family has drawn considerable interest owing to their potential applications in neutron detectors \cite{Montag15}, photovoltaic cells \cite{Bacewicz88}, and Li-ion batteries \cite{Beleanu11}. For many years, this class of filled tetrahedral semiconductors was not explored in the context of thermoelectricity. In 2006, with the help of rigid band approach and semi-classic Boltzmann theory, Madsen investigated a large number of compounds and predicted LiZnSb as a new potential thermoelectric material, with an estimated \textit{ZT}$\sim$2 at 600 K for lightly doped \textit{n}-type LiZnSb \cite{Madsen06}. 

Since then, the Nowotny-Juza phases have witnessed a surge in the field of thermoelectricity. In response to Madsen's calculations, Toberer experimentally reported a \textit{ZT} $<$ 1 of \textit{p}-doped LiZnSb \cite{Toberer09}. In a recent work, the transport properties of Li-based Nowotny-Juza alloys were studied by a first principles approach \cite{Yadav15}. 
Very recently polytypism -- the presence of stable structures with a different layer-stacking sequence -- was reported for the first time in this class of materials. 
The solution phase method revealed the cubic (\textit{F$\bar{4}$3m}) analogue of LiZnSb despite the hexagonal (\textit{P6$_3$mc}) structure reported by traditional solid-state synthesis \cite{White16}. However, the solution phase attempt on LiZnP resulted in the cubic structure in accordance with the solid-state techniques \cite{White16Chem}. Thus, merely varying the synthetic technique may not be solely responsible for the polytypism in Nowotny-Juza phases.

The phenomenon of polytypism is, of course, not limited to Nowotny-Juza phases. The first investigations on polytypism were performed on SiC in 1912. Since then, the polytypism has been reported in a large number of systems. The phenomenon is governed by a number of parameters but in any case greatly influenced by temperature and pressure. The cubic sphalerite and hexagonal wurtzite are two well-known examples of polytypism in ZnS. A number of perovskite-related structures are reported to be polytypic under high pressure \cite{Trigunayat91}.  

Concerning the polytypism in Nowotny-Juza phases, there are quite few open questions. First of all, what is the driving force of polytypism in LiZnSb? Is it achievable by varying the synthetic techniques only or is the transformation purely pressure driven? Is it restricted to \textit{F$\bar{4}$3m} and \textit{P6$_3$mc} or also possible in closely related \textit{P6$_3$/mmc} symmetry? Will transport properties be affected by the phase transition? And most importantly, is it limited only to LiZnSb or also possible in other members of LiZn\textit{X} family? To answer these fundamental questions at least in part, we perform a set of detailed {\it ab initio} calculations for the LiZn\textit{X} systems. 

The first three members of LiZn\textit{X} (\textit{X} = N, P, As, Sb, and Bi) family crystallize in cubic (\textit{F$\bar{4}$3m}) symmetry \cite{Kuriyama99, Kuriyama88, Kuriyama87} whereas the latter two are reported in hexagonal symmetry (\textit{P6$_3$mc}) \cite{Toberer09, Tiburtius77}. On descending from N to Bi in the family, the symmetry changes from cubic to hexagonal at LiZnSb. The polytypism is also exhibited by LiZnSb. It is quite possible that the onset of polytypism in LiZn\textit{X} family begins at LiZnSb and likely to prevail in succeeding LiZnBi also. By the same analogy, moving up from Bi to N, the symmetry changes from hexagonal to cubic at LiZnAs. We assert that the polytypism may also be exhibited by LiZnAs. Stating that LiZnAs and LiZnBi are very likely to show different phases, the polytypism in other family members cannot be ruled out completely. Now, we look for the driving force for such phase transition.

As we move from N to Bi in LiZn\textit{X} family, the increase in the size of pnictogen atoms leads to the expansion of crystal lattice. The relative expansion of the crystal volume corresponds to the negative internal pressure. Since the phase transition appears towards the bottom of the LiZn\textit{X} series i.e. LiZnSb, the role of internal pressure in polytypism should be considered. Thus, along with controlling the synthesis conditions, the internal pressure seems to be a governing factor in the phase transition. A number of studies have noted a profound effect of internal pressure in controlling the phase transitions \cite{Huon17, Horiuchi03, Fratini08}. In addition, hydrostatic pressure is well-known to govern the phase transition in a number of compounds \cite{Bi11, Cao17, Jiang14}. However, there is no previous indication of hydrostatic pressure driven polytypism in Nowotny-Juza phases, but {\it a priori} one should consider hydrostatic pressure also as a potential driving force of polytypism in Nowotony-Juza phases.

Here, we aim to elucidate the physics behind the polytypism behavior in the LiZn\textit{X} (X = N, P, As, Sb, and Bi) family, the role of internal and hydrostatic pressure in controlling such behavior, and, in the end, the effect of polytypism on thermoelectric properties. We also consider the possibility of doping in order to tune the systems towards desirable thermoelectric properties. The paper is organized as follows: Section II consists of a brief description of computational tools utilized. In Sec. III, we discuss the results wherein we describe the crystal structure, structural optimization, phonon stability of LiZn\textit{X} family in cubic and hexagonal symmetry, and electrical and thermal transport properties of LiZn\textit{X} systems in different possible space groups. And finally, we summarize our findings and important conclusions in Sec.~IV.

\section{Computational Details}
We use a combination of two different first-principles density functional theory (DFT) codes: the full-potential linear augmented plane wave method (FLAPW) \cite{Singh06} implemented in Wien2k \cite{Blaha01} and the plane-wave pseudopotential approach implemented in Quantum Espresso package \cite{Giannozzi09}. The former has been used to obtain equilibrium lattice constants, electronic structure, and transport properties, and the latter to confirm the structure stability by determining the phonon spectrum.

The FLAPW calculations are performed using a modified Perdew-Burke-Ernzerhof (PBE correlation) \cite{Perdew08} implementation of the generalized gradient approximation (GGA). For all the calculations, the scalar relativistic approximation is used. In case of LiZnBi, we also performed full relativistic calculations to see the effect of spin-orbit coupling. However, the metallic nature of LiZnBi remained intact even on inclusion of spin-orbit coupling. The muffin-tin radii (RMTs) are taken in the range 1.77--2.5 Bohr radii for all the atoms. RMT {$\times$} kmax = 7 is used as the plane wave cutoff. The self-consistent calculations were employed using 125000 \textit{k}-points in the full Brillouin zone. The energy and charge convergence criterion was set to 10$^{-6}$ Ry and 10$^{-5}$ e, respectively. 

The electrical transport properties have been calculated using the Boltzmann theory \cite{Allen} and relaxation time approximation as implemented in the Boltztrap code \cite{Madsen06Boltztrap}. The Boltztrap code utilizes the input from Wien2k code. The electrical conductivity and power factor are calculated with respect to time relaxation, \textit{$\tau$}; the Seebeck coefficient is independent of \textit{$\tau$}. 

In the plane-wave pseudopotential approach, we use scalar-relativistic, norm-conserving pseudopotentials for a plane-wave cutoff energy of 100~Ry. The exchange-correlation energy functional was evaluated within the GGA, using the Perdew-Burke-Ernzerhof parametrization, and the Brillouin zone is sampled with a 20$\times$20$\times$20 mesh of Monkhorst-Pack \textit{k}-points. The calculations are performed on a 2$\times$2$\times$2 \textit{q}-mesh in the phonon Brillouin zone. 
 
\section{Results}

\subsection{Crystal Structure} 
 
\begin{figure}
\centering
\includegraphics[scale=0.25]{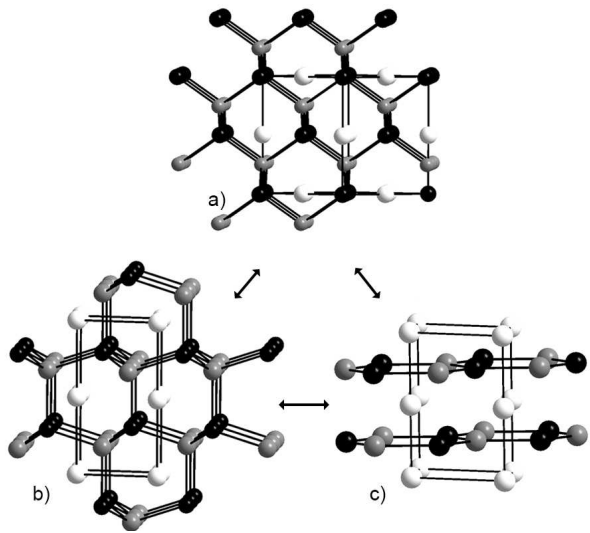}
\caption{The crystal structure of LiZn\textit{X} (\textit{X} = N, P, As, Sb, and Bi) family in three possible geometries a) \textit{F$\bar{4}$3m}, Li at ($\frac{1}{2},\frac{1}{2},\frac{1}{2}$), Zn at origin, and \textit{X} at ($\frac{1}{4},\frac{1}{4},\frac{1}{4}$), b) \textit{P6$_3$mc}, Li at origin, Zn at ($\frac{1}{3},\frac{2}{3}$, \textit{z}), and \textit{X} at ($\frac{2}{3},\frac{1}{3}$, \textit{z$^\prime$}), and c) \textit{P6$_3$/mmc} symmetry, Li at origin, Zn at ($\frac{1}{3},\frac{2}{3}, \frac{1}{4}$), and \textit{X} at ($\frac{1}{3},\frac{2}{3}, \frac{3}{4}$). The white, black, and gray spheres represent Li, Zn, and \textit{X}, respectively. Highlighted the hexagonal framework of Zn and \textit{X} in different symmetries.}
\label{crystal}
 \end{figure}

The first three members of LiZn\textit{X} family viz. LiZnN, LiZnP, LiZnAs, and the recently found cubic LiZnSb crystallize in \textit{F$\bar{4}$3m} symmetry whereas the late members LiZnSb and LiZnBi crystallize in polar \textit{P6$_3$mc} LiGaGe type structure. Another possible symmetry that we consider belongs to nonpolar \textit{P6$_3$/mmc} ZrBeSi type structure (Fig.~\ref{crystal}). The crystal structure in \textit{F$\bar{4}$3m} symmetry can be visualized as binary [Zn\textit{X}]$^-$ in a zincblende type lattice interpenetrated with a face-centered Li$^+$ lattice \cite{Kuriyama99, Kuriyama88, Kuriyama87}. Whereas the polar \textit{P6$_3$mc} structure can be pictured as a wurtzite [Zn\textit{X}]$^-$ lattice stuffed with a Li$^+$ lattice, the nonpolar \textit{P6$_3$/mmc} structure is a planar variant of LiGaGe type structure \cite{Casper08}. 
 
The polar \textit{P6$_3$mc} and nonpolar \textit{P6$_3$/mmc} structures are correlated by a buckling parameter. As can be seen in Fig.~\ref{crystal}, if buckling is eliminated from the hexagonal framework in \textit{P6$_3$mc} symmetry, the resulting structure will be \textit{P6$_3$/mmc} and vice versa. The hexagonal framework is somewhat preserved in the cubic counterpart. However, it may require a great deal of pressure to interconvert the three structures. To understand this in detail, we investigate in the following the structural optimization in different symmetries.

\subsection{Structural Optimization}
The ground state properties of LiZn\textit{X} systems are calculated using the GGA implemented in Wien2k. Fitted with Birch Murnaghan equation \cite{Birch47}, the total energy is minimized as a function of volume, c/a and volume; and c/a, volume, and forces, for \textit{F$\bar{4}$3m}, \textit{P6$_3$/mmc}, and \textit{P6$_3$mc} symmetry, respectively. Table~\ref{optimized_data} lists the calculated cell parameters and band gap values of LiZn\textit{X} systems in different symmetries. The optimized free atomic position along the \textit{z}-axis, in \textit{P6$_3$mc} structure, increases from 0.73 to 0.84 in case of Zn whereas the position of \textit{X} changes from 0.09 to 0.23 on going from LiZnN to LiZnBi. Overall, the distance between Zn and \textit{X} remains almost constant. Further, we tried to establish some known relations between different symmetries i.e. $a_{P6_3mc}$/ $a_{P6_3/mmc}$ $\approx$ 1 and $a_{cubic}$/ $a_{hexagonal}$ $\approx$ $\surd$2. Such correlations are helpful in designing new materials by first-principles calculations for various application purpose. 

The calculated cell parameters of LiZnN, LiZnP, LiZnAs in \textit{F$\bar{4}$3m} and LiZnSb, LiZnBi in \textit{P6$_3$mc} symmetry are in good agreement with reported values. The calculated cell parameters by Wien2k exceeds the solid-state synthesis reported values by 0-0.78\%, except for recently reported low-temperature solution phase cubic LiZnSb (2.8 \%). The unit cell volume, as expected, expands in all the systems in different symmetries with increasing atomic size of \textit{X} element. Besides, the calculated c/a ratio of all but LiZnN in \textit{P6$_3$mc} symmetry is close to the ideal value of 1.633. However, the c/a ratio range from 1.24 to 1.99 in \textit{P6$_3$/mmc} structure. As the symmetry changes from planar \textit{P6$_3$/mmc} to puckered \textit{P6$_3$mc}, excluding LiZnBi, the c/a ratio decreases accordingly. Thus, the introduction of buckling in planar honeycomb lattice results in lowering the c/a ratio. 

\begin{table*}[]
\centering
\setlength{\arrayrulewidth}{0.5pt}
\begin{tabular*}{\textwidth}{c @{\extracolsep{\fill}} cccccccccccccc}
\hline
\hline
                                  & \multicolumn{3}{c}{\textit{F$\bar{4}$3m}}		                        		  &\multicolumn{3}{c}{\textit{P6$_3$/mmc}}                                                     						& \multicolumn{5}{c}{\textit{P6$_3$mc}}                                                                                                      		 \\ \hline
\multicolumn{1}{l|}{System}       & \multicolumn{1}{c}{a (\AA)}     & \multicolumn{1}{c}{E$_g$ (eV)}  	& \multicolumn{1}{c|}{B (GPa)}& \multicolumn{1}{c}{a (\AA)} & \multicolumn{1}{c}{c/a}  & \multicolumn{1}{c}{E$_g$ (eV)}& \multicolumn{1}{c|}{B (GPa)}	& \multicolumn{1}{c}{a (\AA)}       & \multicolumn{1}{c}{c/a}     & \multicolumn{1}{c}{E$_g$ (eV)}  & \multicolumn{1}{c}{B (GPa)}    	& \multicolumn{1}{c}{\textit{z$_{Zn}$}} & \multicolumn{1}{c}{\textit{z$_{X}$}} 	\\ \hline
\multicolumn{1}{l|}{LiZnN}        & \multicolumn{1}{c}{4.92}        & \multicolumn{1}{c}{0.53}       	& \multicolumn{1}{c|}{113.79}	& \multicolumn{1}{c}{3.39}    & \multicolumn{1}{c}{1.99} & \multicolumn{1}{c}{0.40}   	& \multicolumn{1}{c|}{50.80}	& \multicolumn{1}{c}{3.40}    & \multicolumn{1}{c}{1.75}    & \multicolumn{1}{c}{0.32}   & \multicolumn{1}{c}{106.41}     & \multicolumn{1}{c}{0.73}  & \multicolumn{1}{c}{0.09}   					\\ 
\multicolumn{1}{l|}{}             & \multicolumn{1}{c}{(4.90$^a$}  & \multicolumn{1}{c}{(1.98)$^a$}     & \multicolumn{1}{c|}{-}	& \multicolumn{1}{c}{-}       & \multicolumn{1}{c}{-}    & \multicolumn{1}{c}{-}      	& \multicolumn{1}{c|}{-}	& \multicolumn{1}{c}{-}       & \multicolumn{1}{c}{-}       & \multicolumn{1}{c}{-}      & \multicolumn{1}{c}{-}    	  & \multicolumn{1}{c}{-}     & \multicolumn{1}{c}{-}   					\\ 
\multicolumn{1}{l|}{LiZnP}        & \multicolumn{1}{c}{5.76}        & \multicolumn{1}{c}{1.35}       	& \multicolumn{1}{c|}{64.50}	& \multicolumn{1}{c}{4.01}    & \multicolumn{1}{c}{1.81} & \multicolumn{1}{c}{1.30}	& \multicolumn{1}{c|}{31.42}    & \multicolumn{1}{c}{4.03}    & \multicolumn{1}{c}{1.62}    & \multicolumn{1}{c}{1.19}   & \multicolumn{1}{c}{57.41}      & \multicolumn{1}{c}{0.77}  & \multicolumn{1}{c}{0.16}   					\\ 
\multicolumn{1}{l|}{}             & \multicolumn{1}{c}{(5.76)$^b$}  & \multicolumn{1}{c}{(2.04)$^b$}    & \multicolumn{1}{c|}{-}	& \multicolumn{1}{c}{-}       & \multicolumn{1}{c}{-}    & \multicolumn{1}{c}{-}   	& \multicolumn{1}{c|}{-}        & \multicolumn{1}{c}{-}       & \multicolumn{1}{c}{-}       & \multicolumn{1}{c}{-}      & \multicolumn{1}{c}{-}    	  & \multicolumn{1}{c}{-}     & \multicolumn{1}{c}{-}   					\\ 
\multicolumn{1}{l|}{LiZnAs}       & \multicolumn{1}{c}{5.97}        & \multicolumn{1}{c}{0.46}       	& \multicolumn{1}{c|}{55.87}	& \multicolumn{1}{c}{4.16}    & \multicolumn{1}{c}{1.78} & \multicolumn{1}{c}{0.57}	& \multicolumn{1}{c|}{26.56}    & \multicolumn{1}{c}{4.18}    & \multicolumn{1}{c}{1.62}    & \multicolumn{1}{c}{0.39}   & \multicolumn{1}{c}{55.02}      & \multicolumn{1}{c}{0.79}  & \multicolumn{1}{c}{0.17}   					\\
\multicolumn{1}{l|}{}             & \multicolumn{1}{c}{(5.94)$^c$}  & \multicolumn{1}{c}{(1.1)$^c$}     & \multicolumn{1}{c|}{-}	& \multicolumn{1}{c}{-}       & \multicolumn{1}{c}{}     & \multicolumn{1}{c}{-}   	& \multicolumn{1}{c|}{-}        & \multicolumn{1}{c}{-}       & \multicolumn{1}{c}{-}       & \multicolumn{1}{c}{-}      & \multicolumn{1}{c}{-}    	  & \multicolumn{1}{c}{-}     & \multicolumn{1}{c}{-}   					\\
\multicolumn{1}{l|}{LiZnSb}       & \multicolumn{1}{c}{6.41}        & \multicolumn{1}{c}{0.54}       	& \multicolumn{1}{c|}{42.46}	& \multicolumn{1}{c}{4.47}    & \multicolumn{1}{c}{1.74} & \multicolumn{1}{c}{0.20}	& \multicolumn{1}{c|}{19.99}    & \multicolumn{1}{c}{4.46}    & \multicolumn{1}{c}{1.62}    & \multicolumn{1}{c}{0.37}   & \multicolumn{1}{c}{46.54}      & \multicolumn{1}{c}{0.83}  & \multicolumn{1}{c}{0.21}  					\\ 
\multicolumn{1}{l|}{}             & \multicolumn{1}{c}{(6.23)$^d$}  & \multicolumn{1}{c}{-}        	& \multicolumn{1}{c|}{-}	& \multicolumn{1}{c}{-}       & \multicolumn{1}{c}{-}    & \multicolumn{1}{c}{-}   	& \multicolumn{1}{c|}{-}        & \multicolumn{1}{c}{(4.42)$^e$}  & \multicolumn{1}{c}{(1.61)}  & \multicolumn{1}{c}{-}      & \multicolumn{1}{c}{-}   	  & \multicolumn{1}{c}{-}     & \multicolumn{1}{c}{-}  					\\ 
\multicolumn{1}{l|}{LiZnBi}       & \multicolumn{1}{c}{6.61}        & \multicolumn{1}{c}{0.00}       	& \multicolumn{1}{c|}{35.52}	& \multicolumn{1}{c}{4.95}    & \multicolumn{1}{c}{1.24} & \multicolumn{1}{c}{0.00}	& \multicolumn{1}{c|}{41.38}    & \multicolumn{1}{c}{4.60}    & \multicolumn{1}{c}{1.62}    & \multicolumn{1}{c}{-}      & \multicolumn{1}{c}{38.88}      & \multicolumn{1}{c}{0.84}  & \multicolumn{1}{c}{0.23}   				        \\	
\multicolumn{1}{l|}{}             & \multicolumn{1}{c}{-}           & \multicolumn{1}{c}{-}        	& \multicolumn{1}{c|}{-}	& \multicolumn{1}{c}{-}       & \multicolumn{1}{c}{-}    & \multicolumn{1}{c}{-}   	& \multicolumn{1}{c|}{-}        & \multicolumn{1}{c}{(4.57)$^f$}  & \multicolumn{1}{c}{(1.61)}  & \multicolumn{1}{c}{-}      & \multicolumn{1}{c}{-}    	  & \multicolumn{1}{c}{-}     & \multicolumn{1}{c}{-}   					\\ 
\hline
\hline
\end{tabular*}
\caption{Wien2k calculated cell parameters \textit{a} and \textit{c}, band gap E$_g$, and bulk modulus B, of LiZn\textit{X} (\textit{X} = N, P, As, Sb, and Bi) systems in \textit{F$\bar{4}$3m}, \textit{P6$_3$/mmc}, and \textit{P6$_3$mc} symmetry. The \textit{z$_{Zn}$} and \textit{z$_{X}$} are the optimized free positions of Zn and \textit{X}, respectively. The experimentally reported values are listed in parentheses.
$^a$Ref. \cite{Kuriyama99},
$^b$Ref. \cite{Kuriyama88},
$^c$Ref. \cite{Kuriyama87},
$^d$Ref. \cite{White16},
$^e$Ref. \cite{Toberer09},
$^f$Ref. \cite{Tiburtius77}}
\label{optimized_data}
\end{table*}

The deviation of c/a ratio from ideal value in \textit{P6$_3$/mmc} symmetry invalidates the possibility of the existence of LiZn\textit{X} family in hexagonal planar structure. Similarly, the existence of LiZnN in polar \textit{P6$_3$mc} structure is uncertain. Importantly, the ideal c/a ratio is preserved in LiZnP and LiZnAs in \textit{P6$_3$mc} structure, indicating the possible stability of two systems in the hexagonal variant. Thus, LiZnP and LiZnAs may show polytypism among the LiZn\textit{X} family which is one of the focus of our work. Further, to study the effect of polytypism on transport properties, it is important to see how the band gap varies in different symmetries. Let us now look at the second part of Table~\ref{optimized_data} where we have shown the trend of band gap values in different symmetries.  

As GGA is known to underestimate the band gap, the calculated values are lower than the experimental ones. However, the values are consistent with previously calculated values \cite{Kacimi14}. The band gap ranges 0.20--1.35 eV, tailor-made for thermoelectric materials. Notably, among the LiZn\textit{X} systems, LiZnP possess the highest band gap in all three symmetries. The band gap in case of cubic LiZnSb is slightly higher than our expectations. Ignoring this, the band gap of LiZn\textit{X} systems in all three symmetries decreases from LiZnP to LiZnBi with increasing atomic size of \textit{X}. The smaller band gap of LiZnN in comparison to LiZnP has been assessed in detail by Kalarasse and Bennecer and was attributed to the strong \textit{p}-\textit{d} repulsion in LiZnN. The repulsion leads to an upward shift of valence band maximum in energy \cite{Kalarasse06}.
  
Interestingly, except for LiZnBi, the band gap survives on going from cubic to hexagonal or hexagonal to cubic symmetry. Provided semiconductors are the best choices for thermoelectric materials, the LiZn\textit{X} systems in different geometries could be interesting TE prospectives. The predicted stability of LiZnP and LiZnAs in hexagonal structure and the survival of band gap in different symmetries collectively motivates us to see whether the polytypism behavior in LiZn\textit{X} family is possible or not. Hence, it becomes increasingly important to see the energy profile in order to affirm the so far proffered polytypism in LiZn\textit{X} family. And most significantly, what could be the realistic volume and pressure requirements for such behavior.  

\begin{figure}
\centering
\includegraphics[scale=0.42]{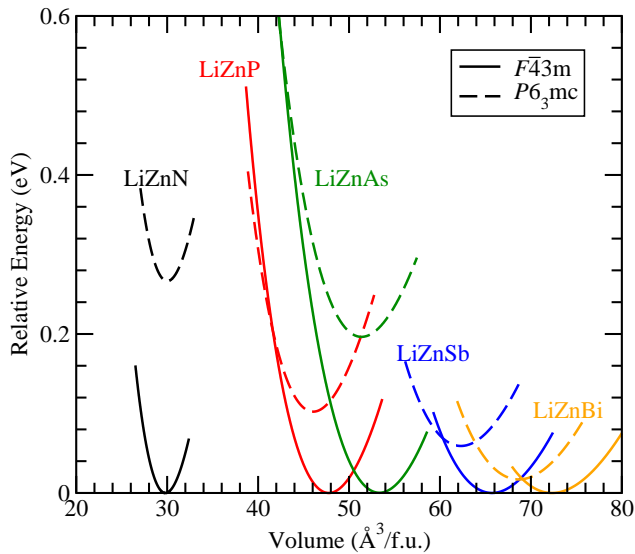}
\caption{Calculated relative energy of LiZn\textit{X} (\textit{X} =  N, P, As, Sb, and Bi) systems as a function of volume per formula unit in cubic \textit{F$\bar{4}$3m} and hexagonal \textit{P6$_3$mc} symmetry. The solid and dashed lines represent cubic and hexagonal symmetry, respectively.}
\label{optimize}
\end{figure}

Figure~\ref{optimize} details the optimization of LiZn\textit{X} systems in \textit{F$\bar{4}$3m} and \textit{P6$_3$mc} symmetry. Initially, we purported the possible existence of LiZn\textit{X} systems in \textit{F$\bar{4}$3m}, \textit{P6$_3$/mmc}, and \textit{P6$_3$mc} symmetry. However, as stated before, the deviations in c/a ratio from ideal value in \textit{P6$_3$/mmc} symmetry indicates their instability in hexagonal planar structure. Furthermore, our calculations reveal that the energy difference between \textit{P6$_3$/mmc} and other symmetries is quite high to achieve under ambient conditions (Table~\ref{energy_data}). Therefore, we rule out any possibility of the existence of LiZn\textit{X} systems in \textit{P6$_3$/mmc} symmetry and proceed with \textit{F$\bar{4}$3m} and \textit{P6$_3$mc} structures in the quest of polytypism. 
 
It is interesting to note that all LiZn\textit{X} systems are found to be stable in the cubic ground state (Fig.~\ref{optimize}). The calculated low energy cubic phases of LiZnN, LiZnP, and LiZnAs are consistent with reported structures \cite{Kuriyama99, Kuriyama88, Kuriyama87}. However, the calculated cubic ground state of LiZnSb and LiZnBi is in contrast to solid-state synthesis reported hexagonal structures (\textit{P6$_3$mc}) \cite{Toberer09, Tiburtius77}. More recently, LiZnSb was synthesized in cubic phase by low-temperature solution phase method \cite{White16}. The same group also showed that the cubic phase is more energetically favorable through first-principles calculations. Here, we would like to raise an important question: Why thermodynamically less stable hexagonal LiZnSb phase was reported for so long despite the recently synthesized more stable cubic LiZnSb? 

In order to understand this, we systematically analyze the energy profile of LiZn\textit{X} systems in both \textit{F$\bar{4}$3m} and \textit{P6$_3$mc} symmetries. Figure~\ref{optimize}, barring LiZnAs, contend the decreasing energy gap between the two phases (\textit{F$\bar{4}$3m} and \textit{P6$_3$mc}) on descending from LiZnN to LiZnBi. The exact values are listed in Table~\ref{energy_data}. The LiZnAs system has a somewhat higher energy gap between the two symmetries than the preceding LiZnP. The decreasing energy gap between \textit{F$\bar{4}$3m} and \textit{P6$_3$mc} symmetry can be attributed to the volume expansion of unit cell on going from LiZnN to LiZnBi. The energy gap between cubic and hexagonal symmetry in LiZnN is 0.48 eV, in LiZnP is 0.30 eV, whereas in case of LiZnAs is 0.77 eV. 
 
However, remarkably, the energy gap between the two symmetries comes down to 0.05 eV for LiZnSb and further down to 0.01 eV in case of LiZnBi. Considering the thermal energy at room temperature is 0.025 eV, a small amount of activation energy by temperature or pressure could interchangeably result in either cubic or hexagonal LiZnSb. This could be the plausible rationale for the synthesis of cubic LiZnSb by low-temperature solution phase method and hexagonal variant by high-temperature solid-state technique. However, as mentioned by White, the thermodynamically less stable hexagonal phase formation by high-temperature solid-state technique is surprising \cite{White16}. 

Nevertheless, in the same line of argument, the even lower energy gap of 0.01 eV between the cubic and hexagonal analogue and the calculated low energy cubic phase of LiZnBi suggests the possible existence of previously known hexagonal LiZnBi in cubic phase. The experimental support of cubic LiZnSb further substantiates our predicted result. Keeping in mind the recent success in cubic LiZnSb, we strongly believe that the cubic analogue of LiZnBi could be synthesized by low-temperature solution phase method. Having attributed that LiZnSb and LiZnBi are most likely to exist in cubic phase complementing the previously reported hexagonal phase, we further investigate the other way round possibility i.e. the existence of previously known cubic phases in the hexagonal phases.

\begin{table}[]
\centering
\setlength{\arrayrulewidth}{0.5pt}
\begin{tabular*}{\columnwidth}{c @{\extracolsep{\fill}} cccccc}
\hline
\hline
System & E$_\textit{F$\bar{4}$3m}$ & E$_\textit{P6$_3$/mmc}$ & E$_\textit{P6$_3$mc}$  &$\Delta$E & P	\\ \hline
LiZnN  & 0                     & 0.48                & 0.26               & -        & -	\\
LiZnP  & 0                     & 0.30                & 0.10               & 0.20     & 12.3 	\\
LiZnAs & 0                     & 0.77                & 0.19               & 0.56     & 20.3	\\
LiZnSb & 0                     & 1.10                & 0.05               & 0.06     & 4.3	\\ 
LiZnBi & 0                     & 0.68                & 0.01               & 0.01     & 1.7	\\
\hline
\hline
\end{tabular*}
\caption{Calculated relative energy, E (in eV) of LiZn\textit{X} systems in \textit{F$\bar{4}$3m}, \textit{P6$_3$/mmc}, and \textit{P6$_3$mc} symmetry. The cubic ground state minimum is set at 0. $\Delta$E (in eV) and P (in GPa) represents the energy and pressure required for the phase transition from cubic to hexagonal symmetry.}
\label{energy_data}
\end{table}

Can hexagonal analogues of previously reported cubic LiZnN, LiZnP, and LiZnAs be synthesized by low-temperature solution phase method? Contrary to expectations, the solution phase synthesis of LiZnP by White et al. also resulted in the cubic structure \cite{White16Chem}. Therefore, merely varying the synthesis technique may not result into polytypism. Once again, the plausible reasoning stems from the energy profile of LiZnP. The energy gap of 0.10 eV between the cubic and hexagonal phase might be responsible for the non-existence of LiZnP in hexagonal symmetry. The energy barrier is even higher for LiZnN (0.26 eV) and LiZnAs (0.19 eV). Thus, further experimental investigations are needed in order to overcome the energy barrier between the cubic and hexagonal symmetry.
  
Few potential approaches utilized over the years involves the application of high-temperature, high pressure or combination of both to achieve the phase transition between different symmetries. In this endeavor, we explore the possibility of phase transition between the well established cubic phase and proposed hexagonal phase of LiZnN, LiZnP, and LiZnAs. We further investigate the cubic to hexagonal phase transition in LiZnSb and LiZnBi.
  
With no possible phase transition in LiZnN, the LiZnP is marked by a cubic to hexagonal phase transition at 0.20 eV whereas the required energy for phase transition in LiZnAs is 0.56 eV (Fig.~\ref{optimize}). Obviously, this may not be achieved under ambient conditions but could be driven by high-temperature or high pressure. Since our calculations are performed at 0 K, the required pressure for cubic to hexagonal phase transition in LiZnP and LiZnAs is calculated to be 12.3 and 20.3 GPa, respectively. In case of LiZnSb and LiZnBi, the required pressure is 4.3 and 1.7 GPa, respectively, corresponding to the 0.06 and 0.01 eV energy requirement. We believe that on the application of the required amount of pressure, one could achieve the hexagonal analogues of these systems. However, due care must be exercised while performing pressure experiments since the bulk modulus indicates the highly compressible nature of LiZn\textit{X} systems (Table~\ref{optimized_data}). 

\begin{figure}
\centering
\includegraphics[scale=0.30]{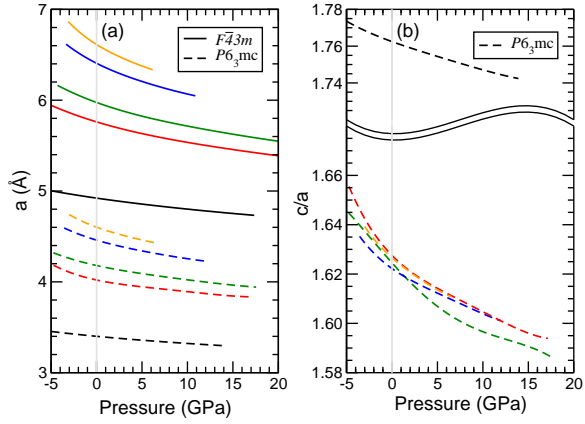}
\caption{(a) Calculated lattice parameter as a function of pressure in cubic \textit{F$\bar{4}$3m} and hexagonal \textit{P6$_3$mc} symmetry and (b) calculated c/a ratio as a function of pressure in hexagonal \textit{P6$_3$mc} symmetry. The solid and dashed lines represent cubic and hexagonal symmetry, respectively.}
\label{lattice}
\end{figure}

As an insight to experimental realization, we further analyze the variation of cell parameter, c/a ratio, and unit cell volume on pressure application. The variation of lattice parameter and c/a as a function of pressure is illustrated in Fig.~\ref{lattice}. The lattice parameter \textit{a} of all the systems decreases steadily in both cubic and hexagonal structures. The decrement of lattice parameter in cubic phase is about 0.02 \AA/GPa in LiZnP and LiZnAs, which grows to 0.04 \AA/GPa in LiZnSb, and steepest in case of LiZnBi, 0.05 \AA/GPa. The decrease in lattice parameter in hexagonal analogues is smaller than their cubic counterpart. The c/a ratio under pressure, except for LiZnN, lies in between 1.62 and 1.58 in the range of 0 to 21 GPa. 
 
\begin{figure}
\centering
\includegraphics[scale=0.42]{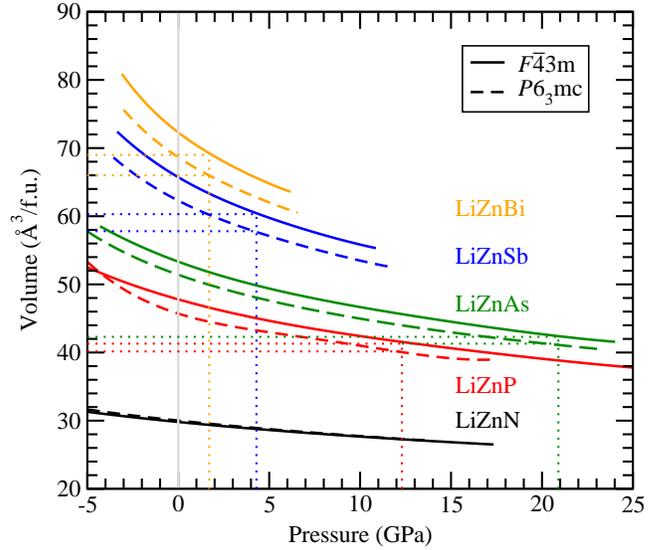}
\caption{Calculated volume per formula unit as a function of pressure in cubic \textit{F$\bar{4}$3m} and hexagonal \textit{P6$_3$mc} symmetry. The solid and dashed lines represent cubic and hexagonal symmetry, respectively. The dotted lines indicate the volume and pressure at cubic to hexagonal phase transition }
\label{volume}
 \end{figure}

Figure~\ref{volume} shows the change in volume under application of pressure. The decrease in volume in cubic LiZnP and LiZnAs is around 0.52 \AA$^3$/GPa, whereas in case of LiZnSb and LiZnBi the decrease in volume is a bit steeper i.e. 1.46 and 1.88 \AA/GPa, respectively. The decrease in volume of hexagonal analogues follows a similar trend i.e. 0.54 \AA$^3$/GPa in LiZnP, 0.51 \AA$^3$/GPa in LiZnAs, 1.44 \AA$^3$/GPa in LiZnSb, and 2.38 \AA$^3$/GPa in case of LiZnBi. The pressure-volume changes suggest that the LiZn\textit{X} systems are compressible under application of pressure. 
 
Highlighted in Fig.~\ref{volume}, the cubic phase stability of LiZnP up to 12.3 GPa, LiZnAs up to 20.9 GPa whereas the LiZnSb and LiZnBi are stable in cubic phase up to 4.3 GPa and 1.7 GPa, respectively. Beyond the specified range of pressure in respective systems, the hexagonal phase is expected to prevail. The corresponding volumes at which cubic to hexagonal phase transition is expected are about 41, 42, 60, and 69 \AA$^3$/f.u. for LiZnP, LiZnAs, LiZnSb, and LiZnBi, respectively. Likewise, the change in volume on cubic to hexagonal phase transition is 1.2, 1.3, 2.5, and 3.5 \AA$^3$/f.u. for LiZnP, LiZnAs, LiZnSb, and LiZnBi, respectively. The discontinuous change in volume in cubic to hexagonal phase transition indicates the transition is of the first order. Notice that our interest is in studying the transport properties of hexagonal variants of cubic ones, therefore, the extreme pressure range at which hexagonal phases may deform is not demonstrated. The idea is to somehow stabilize the cubic systems in hexagonal ones and explore their transport properties. Above all, it seems pertinent to achieve the hexagonal variants of cubic LiZn\textit{X} systems by application of the certain degree of pressure.

As outlined in the introduction, the feasibility of polytypism in LiZn\textit{X} family increases as one moves from LiZnN to LiZnBi i.e. with the relative expansion of the unit cell. The LiZnSb has already been reported as an instance of polytypism  whereas our results indicate a strong possibility of polytypic behavior in LiZnBi. We believe that the relative expansion of the unit cell volume will facilitate the polytypism in early members of LiZn\textit{X} family i.e. LiZnP and LiZnAs whereas it may induce polytypism in LiZnN. By increasing the unit cell volume, one can tone down the conditions required to carry out the phase transition. Therefore, we would like to bring another perspective to manifest the polytypism in LiZn\textit{X} systems: the role of internal pressure. The effect of internal pressure is well-known in inducing phase transition and controlling electronic and magnetic properties in a number of systems. Some of the works on internal pressure can be found in the literature \cite{Huon17, Horiuchi03, Fratini08, Dhital17}. 
 
The internal pressure in a crystal lattice can be achieved by doping with suitable dopants, introducing the nonreactive entities such as hydrogen or helium in the vacant sites or by epitaxial strain in thin films. The doping may result in either positive or negative internal pressure depending upon the choice of dopant whereas the insertion of chemical entities results in expansion of unit cell i.e. negative internal pressure. Since our interest lies in the expansion of unit cell volume, the doping of heavier pnictogen atoms in LiZnN, LiZnP or LiZnAs will subsequently expand the crystal lattice and may facilitate the process of polytypism. The choice of pnictogen atoms as dopants also ensures the intact electronic structure.
  
The introduction of nonreactive elements such as hydrogen and helium in the vacant sites builds the negative internal pressure within the system and eventually expands the crystal lattice \cite{Kraiem01} . However, it may be challenging to introduce the hydrogen and helium at the vacant sites. The solution phase chemical route could be a favorable approach to achieve the hydrogen insertion. Overall, either doping or hydrogen insertion should be tuned in a way that does not distort the original crystal lattice. The discussion on pressure application suggests that internal pressure can complement the hydrostatic pressure in realizing the polytypism in LiZn\textit{X} systems.

Thus far, in response to the recently discovered polytypism in LiZnSb, we propose that LiZnBi could be the most promising candidate for polytypism in LiZn\textit{X} (\textit{X} = N, P, As, Sb, and Bi) family. In addition, provided certain conditions such as pressure application may also induce the polytypism in LiZnP and LiZnAs. The two symmetries explored in the context of polytypism are \textit{F$\bar{4}$3m} and \textit{P6$_3$mc}. To validate our findings, it is crucial to demonstrate the stability of hypothesized structures i.e stability of LiZnN, LiZnP, and LiZnAs in \textit{P6$_3$mc} whereas the stability of LiZnSb and LiZnBi in \textit{F$\bar{4}$3m} symmetry. Thuswise, in the next section, we test the dynamic stability of LiZn\textit{X} systems in cubic and hexagonal structures. 

\subsection{Phonon Stability}
The optimized structures were further tested for dynamic stability through a two-step phonon calculation. First, the Wien2k optimized cell parameters were successfully reproduced by Quantum Espresso. Followed by phonon dispersion calculation by utilizing the Quantum Espresso implemented density functional perturbation theory (DFPT). The calculations were performed on a 2x2x2 mesh in the phonon Brillouin zone, and force constants in real space derived from this input are used to interpolate between \textit{q} points and to obtain the continuous branches of the phonon band structure. 
 
Phonons are the normal modes or quantum of vibrations in a crystal and are an indicative of the stability of a system. In order to be dynamically stable, the phonon frequencies of a system should be real and not imaginary \cite{Elliott06, Togo15}. Table~\ref{phonon_data} summarizes the results of phonon calculations. The `+' sign indicates the positive frequencies throughout the Brillouin zone whereas the `--' sign shows the existence of negative frequencies within the system. Thus, the `+' and `--' sign implies the stability and instability, respectively, in the system. 

\begin{table}[]
\centering
\setlength{\arrayrulewidth}{0.5pt}
\begin{tabular*}{\columnwidth}{c @{\extracolsep{\fill}} ccccc}
\hline
\hline
System & \textit{F$\bar{4}$3m} & \textit{P6$_3$/mmc} & \textit{P6$_3$mc}  \\ \hline
LiZnN  & +                     & --                 & +                   \\
LiZnP  & +                     & --                 & +                   \\
LiZnAs & +                     & --                 & +                   \\
LiZnSb & +                     & --                 & +                   \\ 
LiZnBi & +                     & +                  & +                   \\
\hline
\hline
\end{tabular*}
\caption{The phonon frequencies of LiZn\textit{X} family in \textit{F$\bar{4}$3m}, \textit{P6$_3$/mmc}, and  \textit{P6$_3$mc} symmetry. `+' and `--' indicates the positive and negative phonon frequencies, respectively.}
\label{phonon_data}
 \end{table}

Remarkably, all LiZn\textit{X} systems in \textit{F$\bar{4}$3m} and \textit{P6$_3$mc} symmetry have positive phonon frequencies and are stable in nature. However, with LiZnBi as an exception, all the systems in \textit{P6$_3$/mmc} symmetry are predicted to be unstable. As investigated earlier, the relative energy of LiZn\textit{X} systems in \textit{P6$_3$/mmc} symmetry is quite high, phonon calculations further confirm the non-existence of LiZn\textit{X} systems in \textit{P6$_3$/mmc} symmetry. Therefore, we rule out any possibility of polytypism in LiZn\textit{X} family in \textit{P6$_3$/mmc} symmetry. Given that one of our objectives is to study the polytypism in all LiZn\textit{X} systems, the stability of LiZn\textit{X} in both \textit{F$\bar{4}$3m} and \textit{P6$_3$mc} symmetry lends enough support to our purpose. The energy profile and dynamic stability collectively suggest that not only LiZnSb but other members of LiZn\textit{X} family are also probable of polytypism in \textit{F$\bar{4}$3m} and \textit{P6$_3$mc} symmetry. 
 
Before proceeding ahead, we would like to address the questions posed in this study regarding the polytypism in LiZn\textit{X} (\textit{X} = N, P, As, Sb, and Bi) systems. \textit{What is the driving force for polytypism in LiZnSb?} Our calculations suggest that the competitive energies of cubic and hexagonal phases could be the possible reason behind polytypism in LiZnSb, in addition to underlying mechanisms. \textit{Is it achievable by varying the synthetic techniques only or the transformation is pressure driven? Is it limited only to LiZnSb or also possible for other members of LiZn\textit{X} family?} 

White et al. reveal the first instance of cubic LiZnSb by solution phase method despite the previously known hexagonal phase by solid-state technique. However, the same group achieved the cubic phase of LiZnP by solution phase method in accordance with the solid-state synthesis of cubic LiZnP. Thus, merely varying the synthesis technique may not lead to different phases. Through first-principles calculations, we have demonstrated that by application of the certain degree of pressure, the previously known cubic LiZnP and LiZnAs could be realized in hexagonal symmetry. Likewise, the cubic LiZnSb and LiZnBi could be transformed into hexagonal analogues by application of pressure. Note that the cubic LiZnSb and LiZnBi are reported to be stable in this study despite the previously known hexagonal phases. 

\textit{Is it restricted to \textit{F$\bar{4}$3m} and \textit{P6$_3$/mmc} or also possible in closely related \textit{P6$_3$/mmc}
symmetry?} The relative energy of LiZn\textit{X} systems in \textit{P6$_3$/mmc} symmetry is quite high in comparison to 
\textit{F$\bar{4}$3m} and \textit{P6$_3$mc} symmetry. Further, phonon calculations confirm the dynamic instability of LiZn\textit{X}
systems in \textit{P6$_3$/mmc} symmetry. \textit{Will transport properties be affected by the phase transition?} To investigate the 
effect of polytypism on thermoelectric properties which is the focus of our work, we survey the electrical and thermal transport 
properties of LiZn\textit{X} systems in cubic \textit{F$\bar{4}$3m} and hexagonal \textit{F$\bar{4}$3m} symmetries. 

\subsection{Transport Properties}
In this section first, we highlight the optimal doping levels at which the maximum power factor in cubic and hexagonal LiZn\textit{X} systems can be obtained at different temperatures. The prediction of optimal doping enables the experimentalists to aim at specific doping levels while seeking the best thermoelectric composition. Then, we discuss the plausible explanation of the behavior of power factor, Seebeck coefficient, and electrical conductivity in terms of electronic features. For convenience, hereafter we use the terms PF for power factor, \textit{c}-LiZn\textit{X} for cubic, \textit{h}-LiZn\textit{X} for hexagonal, \textit{p}-LiZn\textit{X} for hole-doped, and \textit{n}-LiZn\textit{X} for electron-doped LiZn\textit{X} systems.
  
We employed rigid band approximation to calculate the electrical transport properties of LiZn\textit{X} (\textit{X} = N, P, As, and Sb) systems. Within this approximation, a single band calculation is sufficient for calculating the transport coefficients at varying doping levels. The approach is effective at low doping levels and has been successfully used in predicting the transport properties of a number of systems \cite{Madsen06Boltztrap, Lee11, Chaput05, Jodin04}. We also assume the constant relaxation time approach which considers the Seebeck coefficient to be independent of relaxation time \textit{$\tau$}, whereas the electrical conductivity and power factor are presented by using \textit{$\tau$} = 10$^{-14}$ s. 

The constant relaxation time approach has been a useful paradigm for predicting the thermoelectric properties through \textit{ab initio} approach. Madsen, in his work, utilized \textit{$\tau$} = 2 x 10$^{-14}$ s while exploring the thermoelectric potential of LiZnSb \cite{Madsen06} which was later found in agreement by Toberer and group \cite{Toberer09}. Toberer et al., with the help of Drude model of conductivity and experimentally determined values for the mobility and effective mass, used an average value, \textit{$\tau$} = 10$^{-14}$ s, for \textit{ZT} estimates of \textit{p}-LiZnSb. In a theoretical investigation on \textit{c}-LiZn\textit{X} systems, a relaxation time of \textit{$\tau$} = 10$^{-14}$ s was used for predicting PF values \cite{Yadav15}. Recently, White et al. used $\kappa_l$/$\tau$ = 1 x 10$^{14}$ for calculating the figure of merit of \textit{c}-LiZnSb \cite{White16}. Overall, in LiZn\textit{X} systems, the relaxation time value of $\tau$ = 10$^{-14}$ has been quite an appropriate figure to predict the transport properties. Therefore, our select of \textit{$\tau$} is quite a rationale which is based on previous theoretical and experimental works. 

Generally, the Seebeck coefficient is maximum when the Fermi level is near the middle of the band gap and decreases as the Fermi level deviates either into valence band or conduction band with doping. Unlikely, the electrical conductivity increases with increasing density of states on either electron or hole doping. The combined effect results in a maximum PF somewhere near the band edge. This is what we have observed for LiZn\textit{X} systems in both cubic and hexagonal symmetries. 

\begin{figure}
\centering
\includegraphics[scale=0.42]{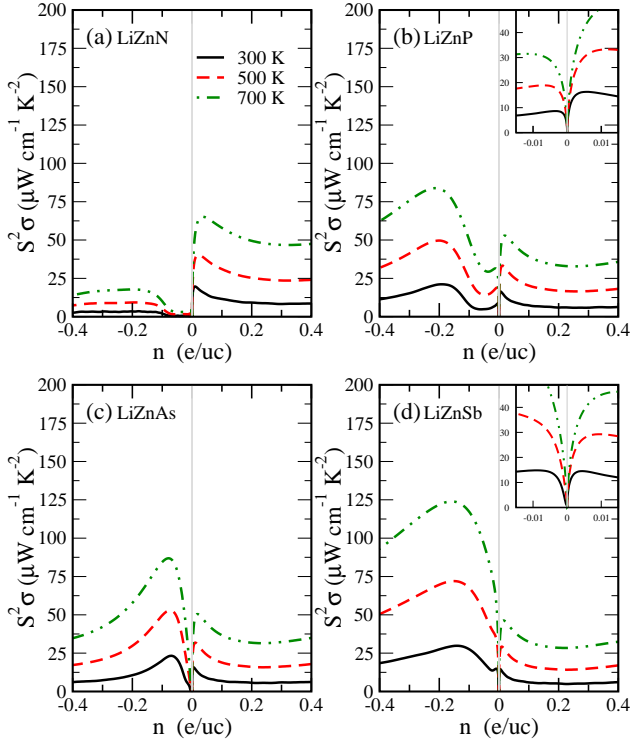}
\caption{Calculated power factor as a function of doping per unit cell in LiZn\textit{X} (\textit{X} = N, P, As, Sb) systems at 300, 500, and 700 K, in \textit{F$\bar{4}$3m} symmetry, assuming $\tau$ = 10$^{-14}$ s. The `+' and `--' sign on horizontal axes respresents hole and electron doping, respectively}
\label{pfcubic}
\end{figure}

\begin{figure}
\centering
\includegraphics[scale=0.42]{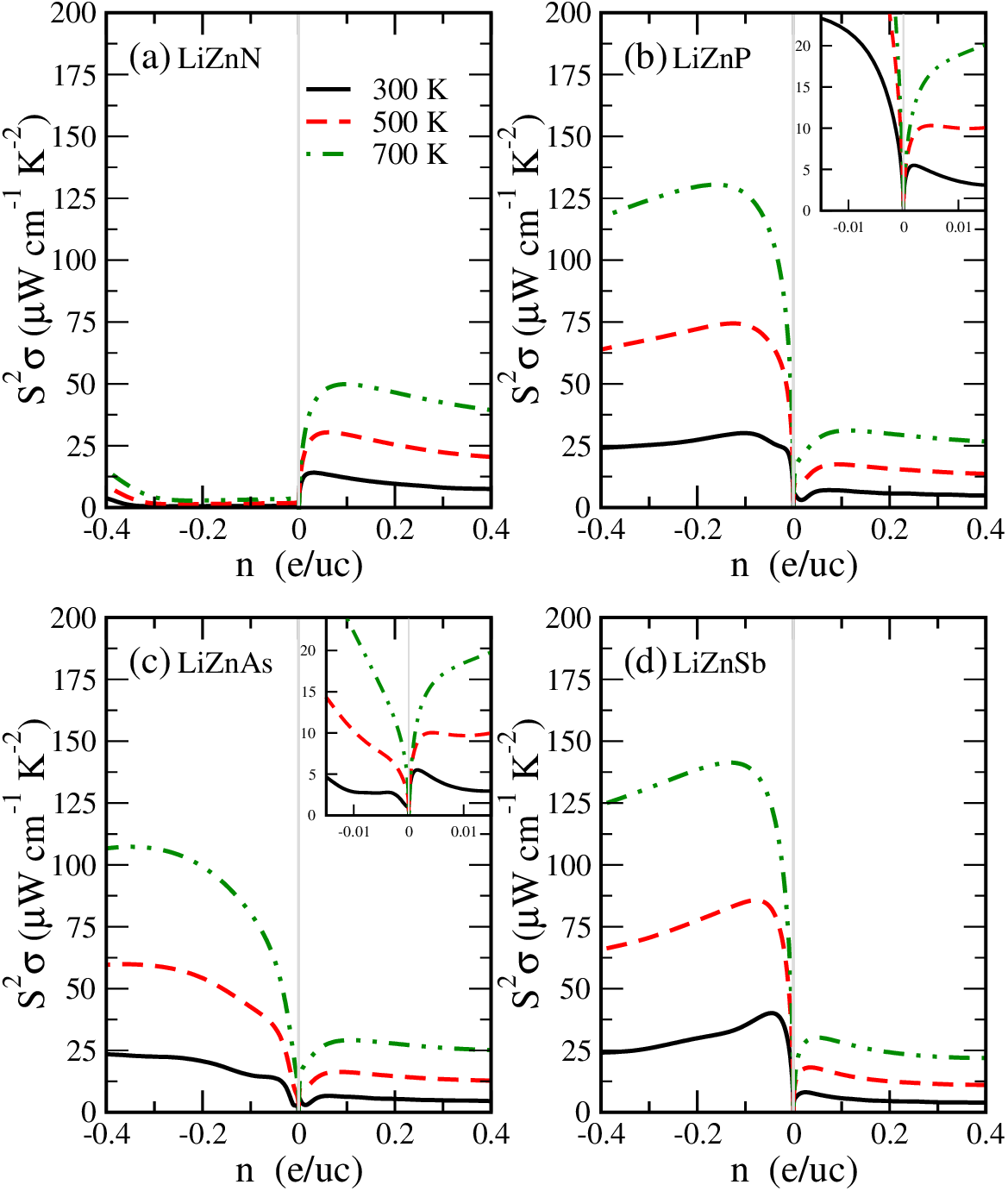}
\caption{Calculated power factor as a function of doping per unit cell in LiZn\textit{X} (\textit{X} = N, P, As, Sb) systems at 300, 500, and 700 K, in \textit{P6$_3$mc} symmetry, assuming $\tau$ = 10$^{-14}$ s. The `+' and `--' sign on horizontal axes respresents hole and electron doping, respectively.}
\label{pfhexa}
\end{figure}

As illustrated in Figs.~\ref{pfcubic} and ~\ref{pfhexa}, the maximum PF on either hole or electron doping is observed near the band edge in LiZn\textit{X} systems in cubic and hexagonal symmetry, respectively. At higher doping levels, the PF falls gradually as the Seebeck coefficient gets significantly reduced. We present our data on PF and doping levels in light of experimental evidence on few LiZn\textit{X} systems. Recently, polycrystalline samples of LiZn$_{1-x}$As were successfully synthesized with less than 15 \% doping by solid-state reaction \cite{Chen16}. The DTA measurements have shown that the melting temperature of LiZnP \cite{Kuriyama91} and LiZnAs \cite{Kuriyama96} are 850 and 950$^0$ C, respectively. 
 
Optimistic of the findings, we present the PF values at room temperature 300 K, 500 K, and as high as 700K while the doping levels are considered up to 0.2 charge carriers per unit cell (e/uc) i.e. 20 \%. However, for better illustration, we have shown the doping levels up to 0.4 e/uc. Before we analyze case by case at different temperatures in different symmetries, we would like to discuss some trends in PF and doping values of LiZn\textit{X} which are common to both cubic and hexagonal symmetry.

The PF increases on going from 300 K to 700 K at all doping levels, depicting the increasing trend of PF with temperature. In all the cases, the maximum PF is obtained at 700 K. Interestingly, within a particular type of doping i.e \textit{n}-type doping, the hexagonal analogues of LiZn\textit{X} have higher PF than their cubic counterpart whereas in \textit{p}-doped LiZn\textit{X} systems the PF is higher in cubic symmetry. Overall, at all considered temperatures, LiZnN is found to have maximum PF on hole doping whilst rest members of LiZn\textit{X} family have better PF on electron doping in either symmetry. 

The optimal doping levels at which maximum PF is obtained shift toward higher values with increasing temperature i.e one has to introduce more number of charge carriers to achieve maximum PF at a higher temperature. The optimal doping levels for \textit{p}-type doping ranges in between 0.004--0.115 e/uc which seems to be quite reasonable for experimental realization. For \textit{n}-type doping, the optimal doping values are comparatively higher and range from 0.066 to as high as 0.347 e/uc. The relevance and acceptability of some higher doping levels are questioned ahead. Briefing the general features of PF of LiZn\textit{X} systems in cubic and hexagonal symmetry, next we focus on the PF and doping of LiZn\textit{X} systems at different temperatures in different symmetries. 

Figure~\ref{pfcubic} shows the PF as a function of doping of LiZn\textit{X} systems at 300, 500, and 700 K in cubic \textit{F$\bar{4}$3m} symmetry. The optimal doping levels for \textit{p}-LiZnN are 0.01, 0.021, and 0.035 e/uc at 300, 500 and 700 K, respectively. The corresponding values of PF are 19.65, 40.41, and 65.17 $\mu$W cm$^{-1}$ K$^{-2}$. The values of PF on electron doping are not convincing even at high electron doping levels, as explained in next section. Thus, only hole doping is worth considering in \textit{c}-LiZnN. Unlike LiZnN, the electron doping is more favorable in LiZnP. The optimal doping levels on electron doping are 0.189, 0.201, and 0.215 e/uc at 300, 500, and 700 K, respectively, the accompanying PF values are 21.11, 49.64, and 83.80 $\mu$W cm$^{-1}$ K$^{-2}$.

We mentioned before that in light of experimental doping levels we propose our results up to 0.20 e/uc. However, only a couple of doping levels exceeds the defined limit. Since the doping level 0.215 e/uc is not far away from 0.20 e/uc, the PF bears close resemblance at the two doping levels i.e. 83.80 $\mu$W cm$^{-1}$ K$^{-2}$ at 0.215 e/uc and 83.65 $\mu$W cm$^{-1}$ K$^{-2}$ at 0.20 e/uc. Therefore, we include the actual figures in order to provide an exact framework to experimentalists. Moreover, in Fig.~\ref{pfcubic}b, a careful observation of \textit{n}-type doping branch of LiZnP reveals a smaller peak of PF on minimal doping complementing the higher PF peak. The inset of Fig.~\ref{pfcubic}b shows the enlarged view of the same. 

At the smaller peak, the PF values are 8.64, 18.90, and 31.48 $\mu$W cm$^{-1}$ K$^{-2}$ at 0.003, 0.007, and 0.011 e/uc optimal doping levels, respectively, corresponding to 300, 500 and 700 K. The values are not worthy of attention in comparison to the larger peak, however, the PF at 700 K i.e. 31.48 $\mu$W cm$^{-1}$ K$^{-2}$, is still an appreciable number. Acknowledging the experimental challenges, the proposed high optimal doping levels of LiZnP are not of much concern since the minimal doping level may also provide a reasonable PF, if not great. 

Moving on to the next systems LiZnAs and LiZnSb, Fig.~\ref{pfcubic}c and ~\ref{pfcubic}d, the better PF values are obtained on electron doping. The PF values of \textit{n}-LiZnAs are 23.26, 52.98, and 86.86 $\mu$W cm$^{-1}$ K$^{-2}$ at 0.068, 0.072, and 0.077 e/uc doping levels, respectively, corresponding to 300, 500, and 700 K. The recently found polytypic LiZnSb has some interesting features to look upon. The optimal doping levels for \textit{n}-LiZnSb are 0.140, 0.149, and 0.161 e/uc at 300, 500, and 700 K, respectively. The corresponding values of PF are 29.82, 72, and 123.92 $\mu$W cm$^{-1}$ K$^{-2}$. The PF values of LiZnSb are quite impressive and best among the LiZn\textit{X} systems in cubic symmetry. The recent solution phase synthesis of \textit{c}-LiZnSb and impressive values of PF further strengthens our motive of polytypism induced thermoelctric properties
in LiZn\textit{X} systems.

An important observation, similar to LiZnP, is the appearance of a small peak at minimal doping at 300 K. However, unlike LiZnP, the peak disappears at higher temperatures. The inset of Fig.~\ref{pfcubic}b highlights the obscured smaller peak at 300 K on \textit{n}-type doping. The PF at this second peak is 14.83 $\mu$W cm$^{-1}$ K$^{-2}$ at a minimal doping of 0.008 electrons per unit cell. Let us now look at the other part of Fig.~\ref{pfcubic}d, the hole doping side. At 300 K, corresponding to optimal doping level 0.004 e/uc, the value of PF is 14.54 $\mu$W cm$^{-1}$ K$^{-2}$. The similar values of PF on \textit{p}-type and \textit{n}-type doping is the most intriguing correlation. 

For device fabrication, it is highly desirable to have \textit{n}-type and \textit{p}-type branches of similar thermoelectric material or ideally of the same material, which is rare. Usually, the performance of one type is inferior to the other \cite{Dubois_Book}. Thus, \textit{c}-LiZnSb could be used for making both \textit{n}-type and \textit{p}-type legs of a thermoelectric module. Unfortunately, the PF at 300 K for the two legs are not substantial. However, considering the benefits of different legs of the same material, \textit{c}-LiZnSb could be used in room temperature thermoelectric applications requiring less efficiency. 

Turning now to the hexagonal symmetry, the PF as a function of doping for LiZn\textit{X} systems is shown in Fig.~\ref{pfhexa}. Alike cubic analogue, the \textit{h}-LiZnN also has higher values of PF on hole doping. Yet again, there is no significant PF whatsoever on electron doping even at higher doping levels at all temperatures. The optimal doping levels at which maximum PF is obtained for \textit{p}-LiZnN are 0.030, 0.062, and 0.096 e/uc at 300, 500, and 700 K, respectively. The values of PF at optimal doping levels are 14.08, 30.40, and 49.84 $\mu$W cm$^{-1}$ K$^{-2}$.

In case of \textit{h}-LiZnP, the electron doping dominates at all temperatures. The optimal doping levels for maximum PF in \textit{n}-LiZnP are 0.102, 0.126, and 0.164 e/uc at 300, 500, and 700 K, respectively. The maximum value of PF at 300 K is 30.07 which drastically increases to 74.42 and 130.49 $\mu$W cm$^{-1}$ K$^{-2}$ at 500 and 700 K, respectively. Similar to \textit{c}-LiZnP, a small peak of PF appears at minimal doping. However, here it appears on \textit{p}-type doping side contrary to \textit{n}-type in \textit{c}-LiZnP. The inset of Fig.~\ref{pfhexa}b highlights that the small peak at minimal doping which emerges on hole doping side persists till 500 K and disappears thereafter. Since the best values of PF are achievable at \textit{n}-type doping, the low PF value at smaller peak is not worth considering.

A similar trend is observed for \textit{h}-LiZnAs. In addition to dominant peak, a small peak is observed at minimal hole doping at 300 K which survives till 500 K. Unfortunately, yet again, the smaller peak corresponds to the less favorable PF and could not assist the high optimal doping levels of LiZnAs. The optimal doping levels of \textit{n}-LiZnAs at three defined temperatures are beyond 0.30 e/uc. Restricting ourselves to the described limit of 0.20 e/uc, the PF values are not far behind than the peak values. This is because of the steady increase in PF with doping, unlike other cases where the PF peak is relatively steeper. The values of PF are 20.79, 54.80, and 100 $\mu$W cm$^{-1}$ K$^{-2}$ at 300, 500, and 700 K, respectively, at 0.20 e/uc.
  
Now we turn to the most promising candidate thus far, \textit{h}-LiZnSb. What makes LiZnSb particularly interesting is the excellent values of PF in cubic phase and the recent experimental realization in cubic symmetry. Further, Madsen's prediction of \textit{h}-LiZnSb as a potential thermoelectric material drives our interest in its properties. The optimal doping levels for \textit{n}-LiZnSb are 0.046, 0.076, and 0.134 e/uc at 300, 500, and 700 K, respectively. The optimal doping levels at 300 and 500 K are quite acceptable in comparison to other \textit{n}-type hexagonal systems whereas the value at 700 K is within our described limit of doping. The PF values at optimal doping levels are 40.10, 85.73, and 141.32 $\mu$W cm$^{-1}$ K$^{-2}$. As expected, the PF values are tremendous and turn out to be the best in our detailed survey of transport properties of LiZn\textit{X} systems in cubic and hexagonal symmetry. As put forward by Madsen \cite{Madsen06} for \textit{h}-LiZnSb and later by White \cite{White16} et al. for \textit{c}-LiZnSb, our calculated results also indicate the thermoelectric potential of cubic and hexagonal LiZnSb.  

\begin{table*}[]
\centering
\setlength{\arrayrulewidth}{0.5pt}
\begin{tabular*}{\textwidth}{c @{\extracolsep{\fill}} cccc}
\hline
\hline
System  & T (K)	& n (e/uc)  		& S$^2\sigma$ ($\mu$W cm$^{-1}$ K$^{-2}$)  	& ZT		\\	\hline
	&300	& +0.010 (+0.030)	& 19.65 (14.08)	   				& 0.14 (0.10)			\\		
LiZnN	&500	& +0.021 (+0.062)	& 40.41	(30.40)	   				& 0.36 (0.29)			\\			
	&700	& +0.035 (+0.096)	& 65.17 (49.84)		   			& 0.55 (0.90)			\\ \hline
	&300	& -0.189 (-0.102)	& 21.11	(30.07)   				& 0.15 (0.24)			\\
LiZnP	&500	& -0.201 (-0.126)	& 49.64	(74.42)	  				& 0.47 (0.92)			\\			
	&700	& -0.215 (-0.164)	& 83.80 (130.49)		   		& 0.78 (1.96)			\\ \hline
	&300	& -0.068 (-0.20)	& 23.26	(20.79)   				& 0.17 (0.16)			\\
LiZnAs	&500	& -0.072 (-0.20)	& 52.98	(54.80)	   				& 0.50 (0.67)			\\			
	&700	& -0.077 (-0.20)	& 86.86 (100.00)		   		& 0.82 (1.49)			\\ \hline
	&300	& -0.140 (-0.046)	& 29.82	(40.10)  				& 0.22 (0.32)			\\ 
LiZnSb	&500	& -0.149 (-0.076)	& 72.00	(85.73)	   				& 0.98 (1.04)			\\			
	&700	& -0.161 (-0.134)	& 123.92 (141.32)		   		& 1.27 (1.95)			\\
\hline
\hline
\end{tabular*}
\caption{Calculated optimal doping levels n, power factor S$^2\sigma$, and \textit{ZT} of LiZn\textit{X} (\textit{X} = N, P, As, Sb) systems at 300, 500, and 700 K in cubic \textit{F$\bar{4}$3m} symmetry. The values in hexagonal \textit{P6$_3$mc} symmetry are enclosed in parentheses. `+' and `--' signs represent \textit{p}-type and \textit{n}-type doping, respectively. The relaxation time is taken to be $\tau$ = 10$^{-14}$ s.}
\label{doping_data}
\end{table*}

In addition to PF, the carrier concentration and Seebeck coefficient at optimal doping levels ranges 10$^{20}$--10$^{21}$ cm$^{-3}$ and 20--138 $\mu$V K$^{-1}$, respectively, in either symmetry. Having discussed in detail the trend and values of PF as a function of doping in LiZn\textit{X} systems at different temperatures, we put forth some salient features of PF values and the significance of actual numbers. We find that \textit{h}-LiZn\textit{X} systems have higher PF values than the cubic ones at optimal doping levels. Returning to the question posed in the beginning, we can state that the hexagonal polytypes of prevailing cubic LiZnN, LiZnP, and LiZnAs will have improved values on \textit{n}-type doping, provided pressure driven cubic to hexagonal phase transition. Nonetheless, the cubic and hexagonal variants of LiZnSb have equally competitive figures.

As far as actual numbers are concerned, the PF values at 300 K ranges in between 19--29 $\mu$W cm$^{-1}$ K$^{-2}$ for cubic symmetry whereas for hexagonal systems the values lie in between 14--40 $\mu$W cm$^{-1}$ K$^{-2}$. At 500 K, the PF values are much improved and lie in the range of 40--72 and 30--85 $\mu$W cm$^{-1}$ K$^{-2}$ for cubic and hexagonal systems, respectively. However, the PF values at 700 K are highly impressive for both cubic and hexagonal symmetry and lie in an overwhelming range of 65--123 and 49--141 $\mu$W cm$^{-1}$ K$^{-2}$, respectively. 

Further, excluding LiZnN, we would like to add that despite the higher PF values obtained at \textit{n}-type doping, the values of \textit{p}-doped cubic LiZn\textit{X} systems at 700 K are not completely inconsiderable. The PF values of cubic LiZnP, LiZnAs, and LiZnSb corresponding to optimal doping levels 0.115, 0.108, and 0.049 e/uc are 31.14, 29.15, and 30.20 $\mu$W cm$^{-1}$ K$^{-2}$ at 700 K, respectively. The optimal doping levels, as can be seen in Fig.~\ref{pfcubic}, are even lower at 300 K and ranges in between 0.004--0.005 e/uc. To give a complete picture to experimentalists, we feature varying doping levels at varying temperatures and at either hole or electron doping. In case one wish to substitute smaller values of dopants keeping an eye on experimental conditions, he may go for \textit{p}-type doping in \textit{c}-LiZn\textit{X} systems rather than aiming high PF values in \textit{n}-type doping. 

As mentioned in the beginning that Nowotny-Juza phases are a special derivative of half Heusler alloys, thus, a requisite comparison between the two would be worthwhile. Despite the lowest values of PF in LiZn\textit{X} systems at 300 K, the values are competitive enough with conventional CoTiSb based materials \cite{Wu07, Qiu09}. Remarkably, the PF values of LiZn\textit{X} systems at 500 K outclass even the best performing half Heusler alloy, FeNbSb \cite{Yu17}. Further, the numbers of PF at 700 K in LiZn\textit{X} systems are too overwhelming to commensurate with parent half Heusler alloys. Altogether, thus far, the evaluation of transport properties implicate that LiZn\textit{X} systems could be a new potential class of thermoelectric material.

The only predicament could be high thermal conductivity which may restrict the \textit{ZT} values. Specifically, the \textit{c}-LiZn\textit{X} systems may have high values of $\kappa$ on account of diamond-like network between Zn and \textit{X}. Madsen used an estimated value of lattice thermal conductivity, $\kappa_l$ = 2 W m$^{-1}$ K$^{-1}$, up to 600 K to calculate \textit{ZT} of \textit{h}-LiZnSb \cite{Madsen06}. Toberer, at a particular carrier concentration, experimentally found that the $\kappa$ of \textit{h}-LiZnSb varies approximately from 6 to 5 W m$^{-1}$ K$^{-1}$ whereas the $\kappa_l$ is of the order of 4--3 W m$^{-1}$ K$^{-1}$ at 300--525 K \cite{Toberer09}. In another work, $\kappa_l$ of \textit{h}-LiZnSb was measured to be 4.5 W m$^{-1}$ K$^{-1}$ at 300 K \cite{Kishimoto08}. Nevertheless, over the years, the $\kappa$ has seen a progressive decrement in a number of systems by isoelectronic alloying, doping or nanostructuring \cite{Snyder08, Chen13, Heremans08}.

Furthermore, Cahill proposed the theoretical lower limit to the thermal conductivity $\kappa_{min}$ of disordered crystals wherein he established that the $\kappa_{min}$ of Si and Ge are 0.6 and 0.9 W m$^{-1}$ K$^{-1}$ at 300 K \cite{Cahill92}. Keeping in mind the similarity of dominant Zn-\textit{X} framework in LiZn\textit{X} systems with diamond-like Si and Ge structure and little exploration of transport properties of LiZn\textit{X} systems, we extend the analogy to correlate their $\kappa$ values. Hence, we believe that $\kappa$ of LiZn\textit{X} systems can be reduced significantly to meet the defined $\kappa_{min}$. We are aware that our assumption is optimistic but the idea is to provide an intuitive estimate of $\kappa_{min}$ values of LiZn\textit{X} systems. Even $\kappa$ of 1 W m$^{-1}$ K$^{-1}$ would yield excellent figure of merit.

Despite some failures, we are currently investigating the theroretical $\kappa_l$ values of cubic and hexagonal LiZn\textit{X} systems. For this study, we use $\kappa_e$ of the parent LiZn\textit{X} systems calculated by Boltztrap code and an estimated value of $\kappa_l$ = 3.5 W m$^{-1}$ K$^{-1}$. Rather than selecting a random constant number of $\kappa_l$, we exploited the experimentally reported $\kappa_l$ values of \textit{h}-LiZnSb. Toberer \cite{Toberer09} reported $\kappa_l$ $\sim$ 4--3 W m$^{-1}$ K$^{-1}$ in the temperature range 300--525 K whereas Kishimoto \cite{Kishimoto08} reported $\kappa_l$ $\sim$ 4.5 at 300 K. We choose an average value of $\kappa_l$ = 3.5 W m$^{-1}$ K$^{-1}$ for estimating the \textit{ZT} values. We believe this value is not overly optimistic because $\kappa_l$ is known to decrease with temperature. Given that the PF values of our interest are observed at 700 K, the value of $\kappa_l$ is indeed on a higher side.

Incorporating the $\kappa$ values as per the discussion, we obtain some encouraging values of \textit{ZT} for experimentalists, listed in Table~\ref{doping_data}. The \textit{ZT} values of LiZn\textit{X} systems ranges in between 0.10--0.32 at 300 K in cubic and hexagonal symmetry. At 500 K, the \textit{ZT} values are in the range of 0.36--0.98 and 0.29--1.04 for cubic and hexagonal symmetry, respectively. The best \textit{ZT} values, as expected, are obtained at 700 K and ranges 0.55--1.27 in case of cubic symmetry whereas 0.90--1.95 for hexagonal symmetry. The most striking values to emerge out from the Table~\ref{doping_data} are of cubic and hexagonal LiZnSb. The \textit{ZT} values of cubic and hexagonal LiZnSb at 700 K are 1.27 and 1.95, respectively, which are complemented by \textit{h}-LiZnP and \textit{h}-LiZnAs. The \textit{ZT} values of \textit{h}-LiZnP and \textit{h}-LiZnAs at 700 K are 1.96 and 1.49, respectively.

Taken together, our findings suggest a promising scope of LiZn\textit{X} systems in thermoelectric applications. Our investigations in this direction are in progress and are likely to explore more corroboration in favor of polytypism and transport properties. As of now, due to lack of experimental investigations of LiZn\textit{X} systems from the thermoelectric perspective, we compare our findings with previously calculated values. Our prediction of hexagonal \textit{n}-LiZnSb as a potential thermoelectric material is in line with Madsen's prediction. Madsen \cite{Madsen06} proposed \textit{ZT} $\sim$ 2 at 600 K for \textit{h}-LiZnSb whereas our calculated value of \textit{ZT} is 1.95 at 700 K. The carrier concentration values of \textit{p}-LiZnSb are fairly consistent with Toberer's reported values \cite{Toberer09}. 

In case of \textit{c}-LiZn\textit{X} systems, except \textit{n}-LiZnAs and \textit{n}-LiZnSb, our predicted optimal doping levels and PF values at 300 K in cubic symmetry are consistent with a previous theoretical work within low doping levels \cite{Yadav15}. Furthermore, our calculated \textit{ZT} value of cubic \textit{n}-LiZnSb is 1.27 at 700 K in comparison to 1.64 at 600 K proposed by White et al \cite{White16}. The slightly lower projected value of ours can be attributed to the choice of different $\kappa_l$ values. While White et al. assumed $\kappa_l$/$\tau$ = 1 x 10$^{14}$ W m$^{-1}$ K$^{-1}$ s$^{-1}$, taking experimental values of $\kappa_l$ into consideration, we utilized $\kappa_l$ = 3.5 W m$^{-1}$ K$^{-1}$. 
 
In spite of this, we obtained a much lower value of \textit{ZT} in case of cubic \textit{p}-LiZnSb. Whereas White reported \textit{ZT} $\sim$ 1.43 at 600 K, our calculated value is only 0.47 at 700 K. Nevertheless, with little experimental investigations, our results are mostly in agreement with previously calculated values. In the next section, we study the electronic structure of cubic and \textit{h}-LiZn\textit{X} systems to delve deeper into the trend and figures of PF and doping type. 

\subsection{Electronic Structure}

\begin{figure}
\centering
\includegraphics[scale=0.42]{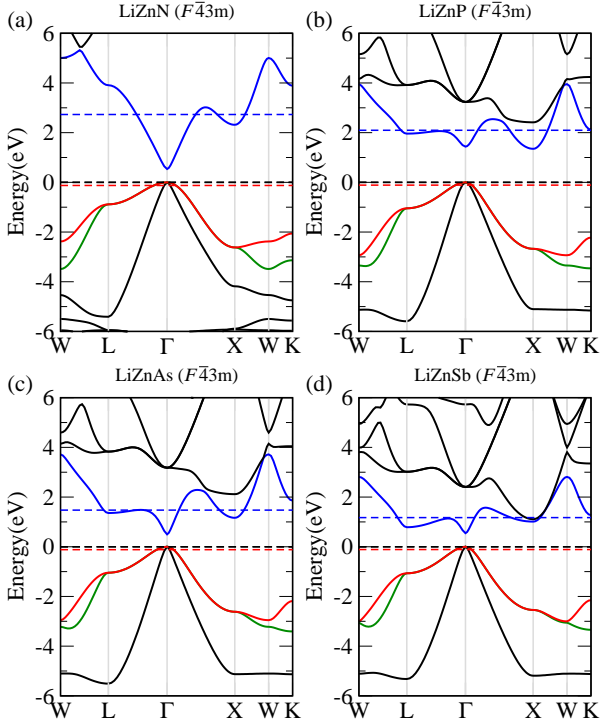}
\caption{Calculated electronic band structures of LiZn\textit{X} (\textit{X} = N, P, As, Sb) systems in \textit{F$\bar{4}$3m} symmetry. The valence band edge is set at zero on the energy axis and represented by a black dashed line. The shift of Fermi level at optimal hole and electron doping at 700 K is highlighted by red and blue dashed line, respectively.}
\label{bscubic}
 \end{figure}

\begin{figure}
\centering
\includegraphics[scale=0.42]{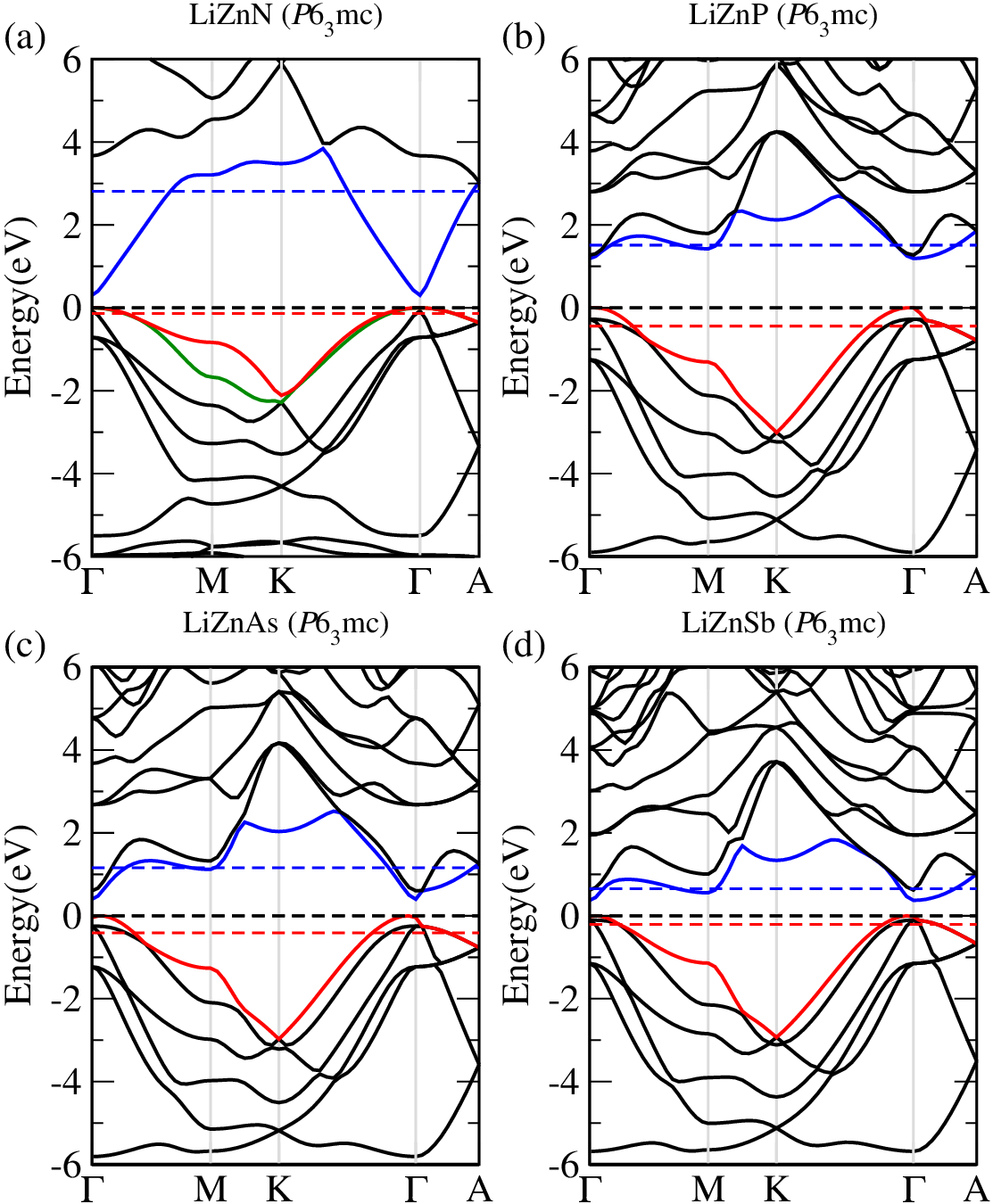}
\caption{Calculated electronic band structures of LiZn\textit{X} (\textit{X} = N, P, As, Sb) systems in \textit{P6$_3$mc} symmetry. The valence band edge is set at zero on the energy axis and represented by a black dashed line. The shift of Fermi level at optimal hole and electron doping at 700 K is highlighted by red and blue dashed line, respectively.}
\label{bshexa}
 \end{figure}

The calculated electronic band structures of LiZn\textit{X} (\textit{X} = N, P, As, and Sb) systems in \textit{F$\bar{4}$3m} and \textit{P6$_3$mc} symmetry are illustrated in Figs.~\ref{bscubic} and ~\ref{bshexa}, respectively. It can be seen, except \textit{c}-LiZnP, the valence band maximum (VBM) and conduction band minimum (CBM) of all the systems are located at $\Gamma$ point in both cubic and hexagonal symmetry, resulting in the direct band gap. In case of \textit{c}-LiZnP, the VBM and CBM are located at different wave vectors $\Gamma$ and X, respectively, making it indirect band gap semiconductor. The optical measurements demonstrated direct band gap of \textit{c}-LiZnN \cite{Kuriyama94} whereas \textit{c}-LiZnP \cite{Kuriyama91} was found to be an indirect band gap semiconductor. However, previous calculations have also reported indirect band gap in case of \textit{c}-LiZnP \cite{Kacimi14}. 

The VBM and CBM edges are dominated by \textit{p}-states of pnictogen atoms and Zn \textit{d}-states (not shown for clarity). The states near VBM have a major contribution from pnictogen atoms whereas states close to CBM are dictated by Zn. There is hardly any contribution from Li in cubic structures, however, Li contributes in states close to CBM in hexagonal symmetry. Such information can be helpful in deciding the dopants in synthesis. For instance, any doping at Li site may not affect VBM in either symmetry whereas doping at Zn and pnictogen site may alter the electronic features. Thus, one can choose the dopants to manipulate the properties as desired. 

Further, the choice of dopant should be cost-effective and more suitably of similar size to the atom to be replaced. The most suitable dopants for \textit{n}-type doping in LiZn\textit{X} systems could be Ga/In/Ge for Zn and Se/Te for Sb. The favorable dopants for \textit{p}-type doping are Co/Fe in place of Zn and Sn/Ge for Sb. 
 
Returning to the band structure, the shift of Fermi level corresponding to optimal doping levels at which maximum PF is obtained at 700 K is highlighted in Figs.~\ref{bscubic} and ~\ref{bshexa}. The red dashed line corresponds to \textit{p}-type whereas blue line to the \textit{n}-type doping. We focus on cubic symmetry first. A common observation among the cubic systems is the threefold degeneracy of VBM. Two bands degenerate in L-$\Gamma$ and $\Gamma$-X direction, the third band is degenerate at $\Gamma$ point, Fig.~\ref{bscubic}. The degenerate bands comprise two relatively heavy and other lighter band which plays an important role in transport properties. The flat and heavy band signifies heavy effective mass of charge carriers which improves Seebeck coefficient whereas the lighter one facilitates the mobility of charge carriers, thereby improving the PF. The bands close to VBM are almost identical in all four systems and it is the CBM which is governing the type of doping. 

Figure ~\ref{bscubic}a, the presence of single parabolic and dispersed band at CBM is an indicative of poor PF on hole doping whereas the CBM of other systems is relatively flat and degenerate at some point with higher bands. This explains the poor PF values on \textit{n}-type doping in \textit{c}-LiZnN as compared to other systems. The combination of heavy and light bands at VBM, as described above, makes \textit{p}-type doping more favorable in \textit{c}-LiZnN. In other systems, the more favored \textit{n}-type doping can be attributed to the dominating features of CBM but the advantages of degenerate heavy and light at VBM cannot be completely ignored. 

The mild optimal \textit{p}-type doping levels in \textit{c}-LiZn\textit{X} systems are attractive from the experimental viewpoint, as noted in transport properties section. Further, the degenerate bands at VBM assure appreciable figures of PF if not significant compared to \textit{n}-type doping. Regarding other systems, the CBM is flat and nondispersive in L-$\Gamma$ direction followed by a more dispersed nature in $\Gamma$-X direction. Here, the single band promotes both Seebeck coefficient and electrical conductivity. The flat band is further assisted by a more dispersed band degenerate somewhere in between X-W direction, depending on the system. Note that this band is degenerate with other higher light bands which influences higher electrical conductivity. Collectively, the PF is improved on \textit{n}-type doping in other \textit{c}-LiZn\textit{X} systems.  

The similar features of CBM of \textit{c}-LiZnP and \textit{c}-LiZnAs are responsible for their similar PF values, 83 and 86 $\mu$W cm$^{-1}$ K$^{-2}$, respectively. The higher PF of \textit{n}-LiZnSb (123 $\mu$W cm$^{-1}$ K$^{-2}$) is because of two factors. First, the CBM of LiZnSb is flatter in L-$\Gamma$ and $\Gamma$-X direction. The contribution of heavy electrons from L and X pocket and lighter electrons from $\Gamma$ pocket maximizes the PF. Secondly, the contribution of electron pocket of the CBM-1 band at X point is more as compared to LiZnP and LiZnAs. 

Similar to cubic symmetry, the VBM of all the systems is more or less same in hexagonal symmetry also. The evaluation of CBM features will be more usable in governing the doping type. In case of \textit{h}-LiZnN, once again the single dispersed band is indicative of low PF on electron doping, as proven in the previous section. The VBM of \textit{h}-LiZnN is twofold degenerate at $\Gamma$. The heavy hole contribution is expected from pockets in $\Gamma$-A region. In rest of all cases, the CBM has a flat band in $\Gamma$-X direction which becomes more dispersed along the direction M-K-$\Gamma$-A. The heavy electrons contribution from flat band region and lighter ones from dispersed band region suggest that \textit{n}-type doping is more favorable. In comparison, the bands at VBM are relatively more dispersed and flat region responsible for better Seebeck coefficient is missing. Except for \textit{h}-LiZnN, the electron pockets at $\Gamma$ and M points are responsible for high PF in \textit{h}-LiZn\textit{X} systems. A description of electronic features of \textit{h}-LiZnSb can be also be found in Madsen's work \cite{Madsen06}. 

\subsection{Discussion and Conclusions}

Materials with different polytypes in many cases exhibit different properties. In a recent major advance, LiZnSb was found to display polytypism and the thermoelectric efficiency of the recently found cubic analogue and hitherto hexagonal LiZnSb were quite extraordinary. In order to understand the driving force behind polytypism and its possible existence in other Li based Nowotny-Juza phases with particular emphasis on thermoelectric properties, we investigated the LiZn\textit{X} (\textit{X} = N, P, As, Sb, and Bi) systems in cubic and hexagonal symmetry.
  
Our detailed theoretical investigations of polytypism in LiZn\textit{X} systems and its effect on transport properties reveal a number of interesting features. In addition to polytypic LiZnSb, LiZnBi turns out to be the most probable system to exhibit the phenomenon. We find that the cubic phase of LiZnBi, similar to LiZnSb, has lower energy in comparison to the existing hexagonal phase. The other cubic members LiZnP and LiZnAs are also likely to undergo cubic to hexagonal phase transition under pressure. Likewise, the cubic LiZnSb and LiZnBi can be pressure driven to respective hexagonal phases. It was found that the pressure driven phase transition could be aided by internal pressure. The combined effect may also bring out polytypism in LiZnN.

As far as transport properties are concerned, the most promising thermoelectric material is LiZnSb. The benefits of LiZnSb are twofold. First, both cubic and hexagonal \textit{n}-LiZnSb have impressive power factors and thus \textit{ZT} values at 700 K. Since the amount of pressure required for the phase transition is not that great (4.3 GPa), there may be chances of a phase transition in realistic application conditions. This is where the significance of high efficiency of the two phases comes into play by ensuring that the performance of the device may not be affected much even when the symmetry changes (from cubic to hexagonal). 

Secondly, as stated by White et al., the constituent elements and low-temperature synthesis make LiZnSb promising from both a cost and a toxicity perspective. Standing on the shoulders of LiZnSb, the power factor and \textit{ZT} values of hexagonal \textit{n}-LiZnP and \textit{n}-LiZnAs are equally promising. The improved values of transport coefficients in hexagonal variants of prevailing cubic LiZnP and LiZnAs are the best instances of polytypism induced thermoelectric performance in LiZn\textit{X} systems, the prime interest of our work.
  
From the theoretical perspective our findings are limited by two factors, the constant relaxation time approximation and estimated lattice thermal conductivity. The relaxation time may vary depending on the actual dopants, temperature, and doping levels. On similar grounds, the values of $\kappa_l$ are expected to change. However, as discussed in the preceding section, the chosen relaxation time and lattice thermal conductivity are in line with previous theoretical and experimental findings. Also, we report conservative estimates -- establishing more precise values is expected to further improve the calculated efficiencies. 
%
%Further, we cannot deny the fact that theoretical predictions are way too easier than experimental realization. To give an illustration, in response to Madsen's prediction of better thermoelectric properties in LiZnSb on \textit{n}-type doping, Toberer's experimental investigation was limited to \textit{p}-type LiZnSb samples. 
%
Experimentalists may further face challenges in synthesizing ordered compositions with controlled doping levels. However, with recent advances in synthetic techniques, we are hopeful of experimental realization of aforementioned LiZn\textit{X} systems with optimal doping levels.

In conclusion, utilizing an \textit{ab initio} approach, we have investigated the occurrence of polytypism and its impact on thermoelctric properties in LiZn\textit{X} (\textit{X} = N, P, As, Sb, and Bi) systems. In addition to LiZnSb, we have found that LiZnBi is most favorable of polytypic behavior whereas the phenomenon could be driven by pressure to attain hexagonal variants of cubic LiZnP and LiZnAs. The improved transport properties on cubic to hexagonal phase transition validates that polytypism may affect better thermoelectric performance. The \textit{ZT} values in cubic and hexagonal LiZnSb at 700 K on \textit{n}-type doping are highly impressive, i.e 1.27 and 1.95, respectively. The other promising values of \textit{ZT} at 700 K are 1.96 and 1.45 of hexagonal LiZnP and LiZnAs, respectively, on \textit{n}-type doping. These findings may serve as a base for experimentalists to explore this new potential class of thermoelectric materials.  

\section{Acknowledgements}
M.Z. is thankful to CSIR for the support of a senior research fellowship. Computations were performed on HP cluster at the Institute Computer Center (ICC), IIT Roorkee, and at IFW Dresden, Germany. We thank Navneet Gupta and Ulrike Nitzsche for technical assistance. H.C.K. gratefully acknowledges financial support from the FIG program of IIT Roorkee (Grant No. CMD/FIG/100596).

\end{document}